\documentclass[letterpaper, 10 pt, conference]{ieeeconf}  

\IEEEoverridecommandlockouts                              

\usepackage{cite}
\usepackage{amsmath,amssymb,amsfonts}
\usepackage{algorithmic}
\usepackage{graphicx}
\usepackage{textcomp}
\usepackage{xcolor}
\usepackage{wrapfig}
\usepackage{multicol}

\title{\LARGE \bf Post-Training Quantization in Brain-Computer Interfaces based on Event-Related Potential Detection}

\author{Hubert Cecotti$^{1}$,  Dalvir Dhaliwal$^{1}$, Hardip Singh$^{1}$,  Yogesh Kumar Meena$^{2}$
\thanks{This work was supported by any organization}
\thanks{$^{1}$HC is with the Department of Computer Science, California State University, Fresno, USA {\tt\small hcecotti@csufresno.edu}}
\thanks{$^{2}$YKM is with the Human-AI Interaction (HAIx) Lab, IIT Gandhinagar, India {\tt\small {yk.meena}@iitgn.ac.in}}%
}

\begin{document}

\maketitle
\thispagestyle{empty}
\pagestyle{empty}

\begin{abstract}
Post-training quantization (PTQ) is a technique used to optimize and reduce the memory footprint and computational requirements of machine learning models. It has been used primarily for neural networks. For Brain-Computer Interfaces (BCI) that are fully portable and usable in various situations, it is necessary to provide approaches that are lightweight for storage and computation. In this paper, we propose the evaluation of post-training quantization on state-of-the-art approaches in brain-computer interfaces and assess their impact on accuracy. We evaluate the performance of the single-trial detection of event-related potentials representing one major BCI paradigm. The area under the receiver operating characteristic curve drops from 0.861 to 0.825 with PTQ when applied on both spatial filters and the classifier, while reducing the size of the model by about $\times$ 15. The results support the conclusion that PTQ can substantially reduce the memory footprint of the models while keeping roughly the same level of accuracy. 
\end{abstract}

\section{Introduction}
\label{sec:introduction}

Non-invasive brain-computer interfaces (BCIs) are accessible to a broader range of users, including individuals with disabilities, children, and elderly individuals~\cite{wolpaw2000brain}. These devices can help individuals with motor impairments communicate, control assistive devices, and interact with their environment more effectively, improving their quality of life and independence~\cite{bockbrader2018brain,chaudhary2016brain}.

Machine learning deployment to the target environment is a challenging task. It is due to the heavy burden of hardware requirements that machine learning, in particular when based on deep learning neural networks~\cite{fang2020post}, models lay on computation capabilities and power consumption~\cite{sze2017hardware,jawandhiya2018hardware}. Methods based on post-training quantification are powerful to accelerate the optimization of such machine learning models.

Post-training quantization (PTQ) is a technique used to optimize and reduce the memory footprint and computational requirements of machine learning models~\cite{hubara2021accurate,banner2019post}. It has been mainly used with large deep-learning architectures. It has advantages that are relevant to non-invasive portable brain-computer interfaces. It is also unknown the extent to which such approaches can impact the classifier's performance.
First, it reduces memory footprint. Quantization reduces the memory required to store model parameters and activations by representing them with lower precision numbers, such as 8-bit integers (int8), instead of 32-bit floating-point numbers. In addition, it can improve inference speed because operations become faster to compute with reduced precision, which leads to improved inference speed. This is especially beneficial for deploying models on resource-constrained devices like mobile phones, IoT devices, or edge devices. This is the case With portable BCIs where the battery is typically located behind the head with limited battery capacity. By decreasing the precision of computations, post-training quantization can lower the power consumption of inference, making it more energy-efficient, which is crucial for battery-powered devices such as portable BCI. Quantized models can be deployed on such hardware, enabling the deployment of advanced machine learning models in situations where it was previously impractical. Other advantages include faster transmission. Quantized models have smaller sizes, making them faster to transmit over networks, which is advantageous for applications requiring frequent model updates or communication with remote servers. In clinical applications that require continuous monitoring of the patient's brain states, it is necessary to exchange data rapidly. Post-training quantization allows for the deployment of deep learning models on cheaper hardware, as the reduced computational and memory requirements enable the use of less powerful processors and smaller memory configurations. This part can be a challenge if the classification is performed on the client side, on a BCI headset, which does not include a GPU for processing the ongoing EEG data. Finally, quantized models are well-suited for deployment on hardware accelerators that are optimized for low-precision computations (i.e., 8-bits).

There are several methods for post-learning weight quantization. In uniform quantization, weights are quantized into a fixed number of levels uniformly across the entire weight range. This is a simple and commonly used method, but it may not always provide the best compression or accuracy. Unlike uniform quantization, non-uniform quantization assigns more bits to regions where the weights are more critical for accuracy and fewer bits to less critical regions. This can potentially improve accuracy compared to uniform quantization. K-Means involves clustering the weight values into a fixed number of centroids and then quantizing the weights to the nearest centroid~\cite{napoleon2010efficient}. It can lead to better compression and accuracy compared to uniform quantization. Despite the reduction in precision, post-training quantization often maintains the model's accuracy within an acceptable range. Advanced techniques such as quantization-aware training can further mitigate any loss in accuracy. With models being used in brain-computer interfaces, it is unknown how the performance will drop or remain steady when using post-learning weight quantization.

The main contribution is the presentation and evaluation of post-training quantization on two classifiers for single-trial detection of event-related potentials. We consider the approach based on the xDAWN spatial filters combined with the BLDA classifier and the Extreme Learning Machine (ELM) classifier.

The paper is organized as follows: The data and methods are provided in Section~\ref{sec:methods}. The performance of the models with and without post-training quantization is given in given in Section~\ref{sec:results}. The impact of the results on portable BCIs is discussed in Section~\ref{sec:discussion}. Finally, the main findings are resumed in Section~\ref{sec:conclusion}.

\section{Methods}
\label{sec:methods}

\subsection{Experimental task}

Nineteen healthy adult participants participated in the study. All the subjects provided written informed consent, reported normal or corrected-to-normal vision, and no history of neurological problems. 

The experimental protocol was reviewed by the California State University, Fresno institutional review board and was following the Helsinki Declaration of 1975, as revised in 2000. Visual stimuli consisted of images ($335 \times 419$ pixel) from the FACES database (2052 images) that was created at the Max Planck Institute for Human Development (MPIB), Berlin, Germany between 2005 and 2007~\cite{ebner2010}. Visual stimuli were centered on the screen (visual angle $\approx 20^o$). Participants were seated comfortably 80~cm from the screen. They were asked to avoid moving during the experiments to avoid muscular artifacts. 

\begin{figure}
\centering
\includegraphics[width=0.8\linewidth]{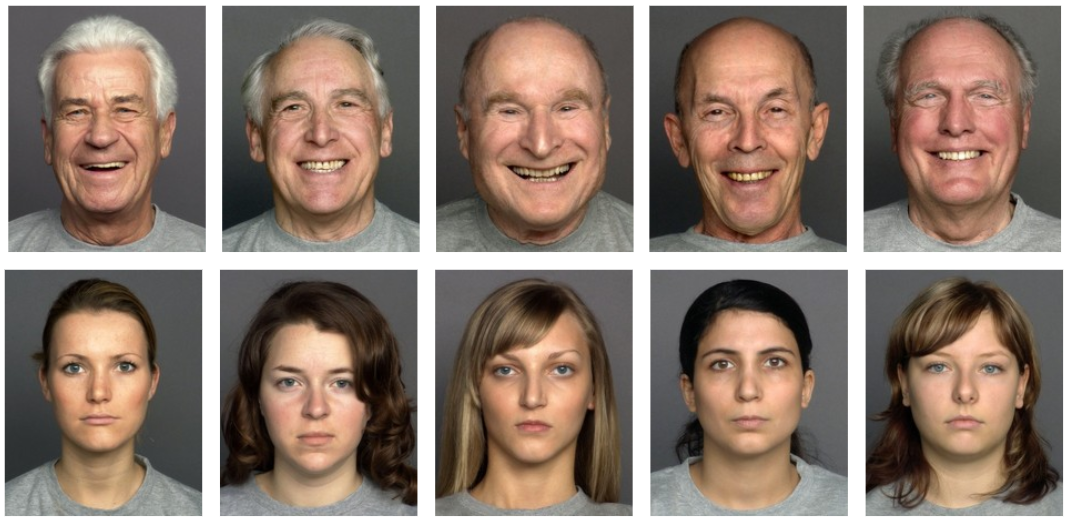}
\caption{Examples of target (top) and non-target (bottom) faces.}
\label{fig:faces}
\end{figure}

The experiment was a rapid serial visual presentation task (presentation of images going at 2~Hz) containing the target and non-target images (target images with a probability of 10\%). Target images are faces of old smiling men, and non-target images are faces of young neutral women (See Fig.~\ref{fig:faces}). 

Each condition has 10 blocks that are separated by a pause, i.e., the participant can start each block when he/she is ready. Each block has 8 sequences, each sequence contains 20 images. With a target probability of 10\%, there are always 2 targets and 18 non-targets in each sequence. It leads to 160 target and 1440 non-target images per condition. 

\begin{figure}
\centering
\includegraphics[width=0.80\linewidth]{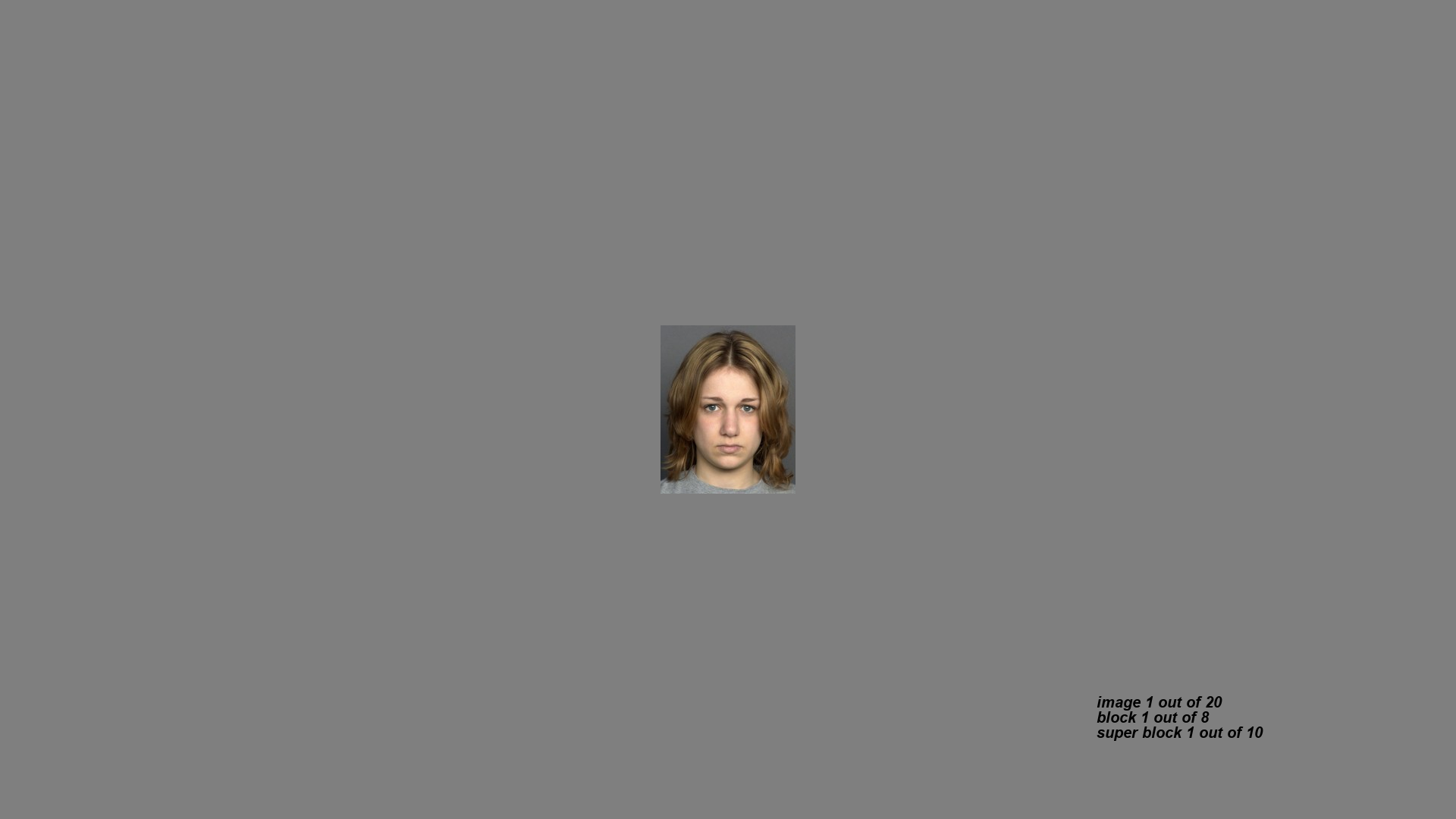}
\caption{Rapid Serial Visual Presentation Task (RSVP) on the computer screen.}
\label{fig:stimulus}
\end{figure}

\subsection{Signal processing and classification}

The EEG signal was acquired with a BioSemi ActiveTwo amplifier with 32 active electrodes placed according to the 10-10 international system. The EEG signal was acquired at 512~Hz then subsequently downsampled to 128~Hz and bandpassed between 0.01 and 42.66~Hz. For single-trial detection, we consider a time segment of 1~second post-stimulus, hence the input is: 128 $\times$ 32.

\subsection{xDAWN+BLDA}

For single-trial detection (target vs. non-target), we consider the combination of the xDAWN spatial filtering approach with the Bayesian Linear Discriminant Analysis classifier (BLDA)~\cite{rivet2009xdawn,rivet2011,hoffmann2008efficient}. This approach has been used in multiple studies for the P300 speller and RSVP tasks. The technique does not require many hyperparameters, so the results can be easily reproduced. The number of spatial filters was set to 8, so the matrix of weights for spatial filtering is 8 $\times$ 32. With the bias, it represents $128*8+1=1025$ weights for the linear classifier.

\subsection{ELM algorithm}

The idea behind the Extreme Learning Machine (ELM) algorithm is that we do not adjust the weights and biases of the input to hidden layer neurons (as we do for normal networks)~\cite{wang2022review,ding2014extreme}. Instead, we randomly assign these weights once and keep them that way until the process is finished. Instead, ELM analytically calculates the output layer weights. This calculation is typically done with the least-square solution.
    $$\beta = (H^TH)^-1*H) * T$$
Where $\beta$ is the output weight.
$(H^TH)^-1*H)$ is the Moore-Penrose inverse of the $H$ matrix (Moore-Penrose inverse is a long series of calculations where a matrix and its transpose are multiplied together. Then we perform the inverse on that and finally multiply it again with the transpose $(H^TH)^-1*H)$). $H$ is the output matrix of the input and hidden layer. \cite{ding2014extreme,wang2022review}.
This is done to minimize the error between the predicted output and the true label. It finds the optimal weight that helps in getting the closest to the desired output.

ELM randomly assigns input layer weights and hidden layer biases, eliminating the need for iterative tuning and bypassing like gradient descent procedures do. After the random assignment, ELM calculates the output layer weights using a simple least squares estimate enabling efficient learning.

ELM learning process can be further represented mathematically as follows.
   $$o_{j} = \sum_{i=0}^H \beta_{i}f(x_{j} * w_{i} + b{i})
    j = 1,...,N$$
where $N$ is the number of training samples, $x_{j}$ = Input, $w_{i}$ are the randomly assigned input weight vector, $b_{i}$ is the randomly assigned hidden layer bias, $f$ is the activation function, $\beta_{i}$ is the analytically calculated output weight vector~\cite{wang2022review,HUANG2006489}.
In the evaluation, we consider 200 hidden units. 

\subsection{Post-weight quantization - ELM}

We consider five conditions for setting the weights between the input and the hidden layer.
\begin{enumerate}
    \item Random weights and biases generated between the range [-1,1].
    \item Input weights and biases initialized to binary values (1 bit)  1 or -1.
    \item Input weighs and biases initialized to binary values (1 bit) 0 or 1.
    \item The weights and biases are set to any of the values (2 bits): [-1,-0.33,0.33,1].
    \item Input weights and biases are initialized to 8 distinct values between -1 and 1. These values have a 2/7 gap: [-1.0000   -0.7143   -0.4286   -0.1429    0.1429    0.4286    0.7143    1.0000].
\end{enumerate}
\vspace{3mm}
We consider two conditions for setting the weights between the hidden layer and the output layer after their evaluation. It is possible to quantize the calculated weights using a histogram (256 bins) by selecting the mean value of each bin. First, a histogram of the weights is created. For the second condition, where we used the weights -1 and 1, it would show two bins (-1 and 1). If we had 10 weights where six were -1 and four of them 1, this will be visually represented in the histogram. We denote by 1, 2, 3, 4, and 5 the five conditions, and 11, 12, 13, 14, 15, the conditions with the histogram quantization of the weights between the hidden layer and the output layer.

\subsection{Post-weight quantization - xDAWN+BLDA}

We consider two types of integers that can be positive and negative. With int4 (4 bits), there is one bit for the sign and 3 bits for the value (from -8 to 7). With int8 (8 bits), there is one bit for the sign and 7 bits for the values ranging from -128 to 127. For the quantization, we consider two solutions: 1) the maximum value of the absolute value of the set of weights to quantize, or 2) the minimum and maximum value of the set of weights to quantize. In the latter case, it is just a linear mapping from a range of values $[v_{min},v_{max}$ to $[-128,127]$ or $[-8,7]$ with only integers. We denote by 1, 2, 3, and 4 the methods: max+int4, max+int8, min-max+int4, min-max+int8.

In the subsequent section, we consider the values that are used for the normalization of the int4 or int8 as doubles (64 bits) but floats (32 bits) could be used as well without a drop in performance. 

\subsection{Performance evaluation}

For the performance evaluation, we consider the area under the receiver operating characteristic curve (AUC) with a five-fold cross-validation. We denote by 0 the absence of weight quantization. Before the ELM classifier, we also consider the xDAWN spatial filters without quantization of the spatial filters.

\section{Results}
\label{sec:results}

Th results are presented in Tables~\ref{tab:results_blda} and ~\ref{tab:results_elm}. They include the AUC performance with the corresponding size (in bits) of the models. The size related to the model is decomposed into the weights of the spatial filters (filter) and the weights of the classifier (classifier). The AUC across subjects is $0.861\pm0.097$ without quantization of the model's parameters with the xDAWN+BLDA classifier. By considering the int4+max for both the weights of the spatial filters and the weights and bias of the classifier, the performance drops to $0.825\pm0.109$. By considering the quantization only for the spatial filters, the best approach reaches 0.864, higher than the default condition. When the quantization is applied to the classifier, the highest performance is observed with int8 combined with the min/max approach. 

The performance with the ELM classifier is lower compared to BLDA, as the mean AUC is only $0.692\pm0.100$ across subjects. By considering the different post-training quantization approaches, it is possible to substantially reduce the size of the model, first by considering only 1 bit for the first hidden layer, and second, by adding the quantization through the histogram, where each weight is one byte, in addition to the normalization parameters that remain doubles.
The performance with the ELM demonstrates the interest of the approach and shows the variability of performance across classifiers, as the BLDA classifier provides better results. 

\begin{table*}[!ht]
    \centering
    \caption{AUC for single-trial detection using the XDAWN+BLDA method}
    \begin{tabular}{l|cccccccccc} \hline
Method &    0/0  & 1/0 & 2/0 & 3/0 & 4/0 & 0/1 & 0/2 & 0/3 & 0/4 & 1/1 \\ 
Filter& 	16384&	1088&	2112&	1152&	2176&	16384&	16384&	16384&	16384&	1088\\
Classifier& 65600&	65600&	65600&	65600&	65600&	8264&	4164&	8328&	4228&	4164\\
Total&	81984&	66688&	67712&	66752&	67776&	24648&	20548&	24712&	20612&	5252\\ \hline				
1&	0.708&	0.664&	0.713&	0.550&	0.696&	0.697&	0.709&	0.704&	0.709&	0.616\\
2&	0.774&	0.776&	0.776&	0.778&	0.777&	0.770&	0.774&	0.777&	0.774&	0.668\\
3&	0.902&	0.899&	0.906&	0.906&	0.906&	0.898&	0.902&	0.897&	0.902&	0.890\\
4&	0.916&	0.916&	0.917&	0.900&	0.914&	0.913&	0.916&	0.915&	0.917&	0.847\\
5&	0.796&	0.796&	0.797&	0.790&	0.796&	0.791&	0.795&	0.794&	0.795&	0.792\\
6&	0.962&	0.963&	0.961&	0.961&	0.962&	0.960&	0.962&	0.961&	0.962&	0.950\\
7&	0.983&	0.983&	0.983&	0.985&	0.984&	0.982&	0.983&	0.983&	0.983&	0.973\\
8&	0.937&	0.930&	0.937&	0.935&	0.937&	0.933&	0.937&	0.936&	0.937&	0.926\\
9&	0.932&	0.922&	0.933&	0.929&	0.931&	0.931&	0.931&	0.932&	0.932&	0.908\\
10&	0.760&	0.758&	0.764&	0.766&	0.762&	0.758&	0.761&	0.756&	0.760&	0.753\\
11&	0.749&	0.733&	0.750&	0.758&	0.750&	0.746&	0.751&	0.743&	0.750&	0.686\\
12&	0.846&	0.839&	0.843&	0.833&	0.840&	0.847&	0.845&	0.836&	0.846&	0.818\\
13&	0.852&	0.852&	0.851&	0.837&	0.852&	0.845&	0.852&	0.851&	0.852&	0.830\\
14&	0.670&	0.685&	0.694&	0.696&	0.694&	0.660&	0.671&	0.674&	0.671&	0.689\\
15&	0.958&	0.954&	0.958&	0.960&	0.959&	0.952&	0.958&	0.954&	0.958&	0.919\\
16&	0.972&	0.973&	0.974&	0.972&	0.974&	0.974&	0.972&	0.970&	0.972&	0.921\\
17&	0.961&	0.962&	0.962&	0.958&	0.961&	0.960&	0.961&	0.961&	0.961&	0.952\\
18&	0.852&	0.846&	0.851&	0.846&	0.851&	0.832&	0.851&	0.818&	0.852&	0.774\\
19&	0.834&	0.803&	0.836&	0.822&	0.839&	0.826&	0.836&	0.826&	0.836&	0.762\\ \hline
Mean&	0.861&	0.855&	0.864&	0.852&	0.862&	0.857&	0.861&	0.857&	0.862&	0.825\\
SD&	0.097&	0.1&	0.093&	0.112&	0.095&	0.099&	0.096&	0.097&	0.096&	0.109\\ \hline
\end{tabular}
\label{tab:results_blda}
\end{table*}

\begin{table*}[!ht]
    \centering
    \caption{AUC for single-trial detection using the XDAWN+ELM method}
    \begin{tabular}{l|cccccccccc} \hline
Method &    1 & 2 & 3 & 4 & 5 & 11 & 12 & 13 & 14 & 15 \\  \hline
1&	0.525&	0.545&	0.511&	0.532&	0.512&	0.552&	0.542&	0.257&	0.554&	0.530\\
2&	0.640&	0.627&	0.509&	0.637&	0.586&	0.582&	0.629&	0.391&	0.645&	0.622\\
3&	0.702&	0.705&	0.517&	0.709&	0.685&	0.662&	0.652&	0.498&	0.712&	0.666\\
4&	0.685&	0.671&	0.477&	0.657&	0.649&	0.699&	0.701&	0.408&	0.709&	0.654\\
5&	0.630&	0.615&	0.465&	0.635&	0.619&	0.608&	0.612&	0.295&	0.630&	0.588\\
6&	0.795&	0.820&	0.492&	0.830&	0.834&	0.812&	0.811&	0.497&	0.815&	0.803\\
7&	0.834&	0.840&	0.531&	0.844&	0.839&	0.841&	0.869&	0.445&	0.827&	0.860\\
8&	0.855&	0.850&	0.501&	0.851&	0.833&	0.855&	0.850&	0.545&	0.821&	0.860\\
9&	0.681&	0.750&	0.492&	0.692&	0.711&	0.691&	0.706&	0.503&	0.721&	0.746\\
10&	0.654&	0.635&	0.520&	0.648&	0.620&	0.636&	0.594&	0.504&	0.620&	0.625\\
11&	0.583&	0.580&	0.476&	0.575&	0.554&	0.568&	0.576&	0.336&	0.585&	0.574\\
12&	0.724&	0.726&	0.440&	0.750&	0.727&	0.678&	0.726&	0.358&	0.718&	0.735\\
13&	0.674&	0.656&	0.449&	0.699&	0.658&	0.707&	0.670&	0.407&	0.663&	0.642\\
14&	0.594&	0.507&	0.477&	0.565&	0.544&	0.522&	0.520&	0.316&	0.560&	0.559\\
15&	0.730&	0.731&	0.504&	0.760&	0.730&	0.698&	0.716&	0.496&	0.723&	0.722\\
16&	0.875&	0.850&	0.492&	0.861&	0.865&	0.858&	0.860&	0.510&	0.859&	0.853\\
17&	0.779&	0.809&	0.520&	0.831&	0.817&	0.798&	0.827&	0.530&	0.827&	0.803\\
18&	0.609&	0.598&	0.463&	0.565&	0.573&	0.592&	0.600&	0.511&	0.623&	0.604\\
19&	0.578&	0.533&	0.417&	0.562&	0.511&	0.554&	0.557&	0.145&	0.549&	0.566\\ \hline
Mean&	0.692&	0.687&	0.487&	0.695&	0.677&	0.680&	0.685&	0.419&	0.693&	0.685\\
SD&	0.100&	0.112&	0.030&	0.111&	0.118&	0.109&	0.114&	0.110&	0.101&	0.11\\ \hline
\end{tabular}
\label{tab:results_elm}
\end{table*}    

By using a Wilcoxon signed-rank test with Bonferroni correction, we provide the results of the pairwise analysis (1 for significant, 0 for not significant) in Tables~\ref{tab:h_blda} and~\ref{tab:h_elm}. The results in Table~\ref{tab:h_blda} indicate that the post-training quantization for both spatial filters and the classifiers provided the worst results, with a drop of about 0.04 in the AUC for the xDAWN+BLDA approach. While there is a drop in performance, it allows to reduce the size of the model by a factor of 15.61. There are also other differences between the baseline and the post-training quantization with condition 0/1.

\begin{table}[!ht]
    \centering
    \caption{Statistical significance through pairwise comparisons (xDAWN+BLDA).}
    \begin{tabular}{@{}l@{}|cccccccccc@{}} \hline
Method &    0/0  & 1/0 & 2/0 & 3/0 & 4/0 & 0/1 & 0/2 & 0/3 & 0/4 & 1/1 \\  \hline
0/0 & x&	0&	0&	0&	0&	1&	0&	0&	0&	1\\
1/0 & -&	x&	1&	0&	1&	0&	0&	0&	0&	1\\
2/0 & -&	-&	x&	0&	0&	1&	0&	1&	0&	1\\
3/0 & -&	-&	-&	x&	0&	0&	0&	0&	0&	1\\
4/0 & -&	-&	-&	-&	x&	1&	0&	0&	0&	1\\
0/1 & -&	-&	-&	-&	-&	x&	1&	0&	1&	1\\
0/2 & -&	-&	-&	-&	-&	-&	x&	0&	0&	1\\
0/3 & -&	-&	-&	-&	-&	-&	-&	x&	0&	1\\
0/4 & -&	-&	-&	-&	-&	-&	-&	-&	x&	1\\ \hline
\end{tabular}
\label{tab:h_blda}
\end{table}

\begin{table}[!ht]
    \centering
    \caption{Statistical significance through pairwise comparisons (xDAWN+ELM).}
    \begin{tabular}{@{}l@{}|cccccccccc@{}} \hline
Method &  1 & 2 & 3 & 4 & 5 & 11 & 12 & 13 & 14 & 15 \\  \hline
1& x&	0&	1&	0&	0&	0&	0&	1&	0&	0\\
2& -&	x&	1&	0&	0&	0&	0&	1&	0&	0\\
3&-&	-&	x&	1&	1&	1&	1&	0&	1&	1\\
4&-&	-&	-&	x&	1&	0&	0&	1&	0&	0\\
5&-&	-&	-&	-&	x&	0&	0&	1&	0&	0\\
11&-&	-&	-&	-&	-&	x&	0&	1&	0&	0\\
12&-&	-&	-&	-&	-&	-&	x&	1&	0&	0\\
13&-&	-&	-&	-&	-&	-&	-&	x&	1&	1\\
14&-&	-&	-&	-&	-&	-&	-&	-&	x&	0\\ \hline
\end{tabular}
\label{tab:h_elm}
\end{table}

\section{Discussion}
\label{sec:discussion}

A key question that was addressed in this paper was whether post-training quantization has an impact on the accuracy of single-trial event-related potential detection. While the answer is positive, the drop in performance is not substantial and does not represent an obstacle to using such an approach in portable BCI applications. In addition to improving BCI accuracy, it is necessary to provide lightweight models that can be deployed on hardware with limited memory and, more importantly, with limited power consumption. 

The development of portable non-invasive BCIs is critical and requires considering the computational cost and memory storage of the models being used. The battery should last long enough to allow the user to have some independence. With brain monitoring devices, EEG signal processing and classification can occur in real-time at a fast pace, hence requiring many decisions over time. 

Traditional BCIs are often bulky and require users to be stationary during operation. Portable BCIs, however, allow users to move around freely while using the device, enabling applications in real-world environments (e.g., home, workplace, or community settings). Furthermore, portable BCIs offer improved user experience by eliminating the need for tethered connections or cumbersome equipment. Users can wear these devices discreetly, leading to greater comfort and acceptance, particularly in social or public settings.

Portable non-invasive BCIs make brain-computer interaction accessible to a broader range of users, including individuals with severe disabilities or mobility impairments. These devices enable users to control computers, prosthetic limbs, or other assistive devices using their brain signals, enhancing their independence and quality of life without being connected to a desktop computer. Portable BCIs have significant potential for clinical applications (e.g., rehabilitation, neurofeedback therapy, and monitoring of neurological conditions). Such devices can be used in home-based rehabilitation programs, remote patient monitoring, and early detection of neurological disorders, leading to improved healthcare outcomes and reduced healthcare costs. 
Finally, such systems can support various applications beyond assistive technologies, including gaming, entertainment, mental health monitoring, and cognitive enhancement. Their portability and ease of use open up diverse opportunities for research, development, and commercialization. 

Future works will deal with the change of the models so the computation is performed on int8, and to determine to what extent it impacts the energy consumption. While it is possible to quantize the model with values on 4 bits,  4-bit computing is mostly obsolete, and present-day programming languages do not support 4-bit data types. 

\section{Conclusion}
\label{sec:conclusion}

Post-training quantization is not reserved for deep learning architecture and can be implemented on different types of classifiers. Post-training quantization represents one step toward the deployment of BCI on hardware with limited resources, where the analysis of brain responses is performed at a high pace with limited resources. While it is necessary to decompose storage requirements from processing requirements, we have shown that lots of information could be stored on 4-bit integers for the spatial filters and/or the model. 

\subsection*{Acknowledgment}
This study was supported by the NIH-R15 NS118581 project.

\bibliographystyle{IEEEtran}
\bibliography{references}

\end{document}